\documentclass[aps,pre,twocolumn,showpacs,groupedaddress]{revtex4}

\usepackage{amssymb,amsmath}
\usepackage{graphicx}

\newcommand{\m}{\cdot}
\newcommand{\ti}{\tilde{i}}
\newcommand{\tb}{\tilde{b}}

\begin{document}

\title{Analytical description of finite size effects for RNA 
secondary structures}

\author{Tsunglin Liu and Ralf Bundschuh}

\affiliation{Department of Physics, Ohio State University, 174 W 18th Av., Columbus OH 43210-1106}

\begin{abstract}
The ensemble of RNA secondary structures of uniform sequences is studied 
analytically. We calculate the partition function for very long sequences 
and discuss how the cross-over length, beyond which asymptotic scaling 
laws apply, depends on thermodynamic parameters. For realistic choices of 
parameters this length can be much longer than natural RNA molecules. This 
has to be taken into account when applying asymptotic theory to interpret 
experiments or numerical results.
\end{abstract}


\maketitle

\section{Introduction}

Folding of biopolymers is a fundamental process in molecular biology without
which life as we know it would not be possible. In biopolymer folding,
well-characterized interactions between individual monomers make a polymer
fold into a specific structure believed to minimize the total interaction free
energy. The apparent simplicity in the formulation of this biopolymer folding
problem is in sharp contrast with the immense challenges faced in actually
describing biopolymer folding quantitatively caused by the intricate interplay
of monomer-monomer interactions and the constraint that the monomers are
connected into a chain of a certain sequence. The biological importance of
biopolymer folding paired with this immense intellectual challenge has sparked
numerous computational and theoretical studies~\cite{profold}. These studies do not
only attempt quantitative predictions of specific structures but also focus on
more fundamental properties of the biopolymer folding problem such as its
phase diagram.

While the bulk of the work concentrates on the folding of proteins due to its
overwhelming importance in pharmaceutical applications, recently RNA folding
has been identified as an ideal model system for biopolymer folding~\cite{tinoco,higgs00}. 
RNA is a biopolymer of four different bases G, C, A, and U. The most important
interaction among these bases is the formation of Watson-Crick(WC) base
pairs, i.e., A--U and G--C pairs. This comparatively simple interaction scheme
makes the RNA folding problem very amenable to theoretical approaches
without losing the overall flavor of the general biopolymer folding problem.
Again, a lot of effort has been devoted to understanding fundamental
properties of RNA folding such as the different thermodynamic phases an
ensemble of RNA molecules can be in as a function of temperature, an external
force acting on the molecules, and the sequence design~\cite{muller,higgs96,pagna,
hart,bund99,degen}.
  
All these theoretical approaches are concerned with the phase behavior of RNA
molecules in the {\em thermodynamic limit}. In order to compare these
theoretical predictions with numerical or actual biological experiments it is
thus important to know which role {\em finite size effects} play, i.e., at
which size of a molecule the universal predictions of the asymptotic theories
are expected to hold.  In this publication we precisely aim to answer this
question. We study homogeneous RNA sequences which allows us to {\em
  analytically} solve for the universal asymptotic behavior as well as the
cross-over length below which the universal theory is not applicable any more.
We find that this cross-over length is very strongly dependent on the sequence
of the molecule. For realistic energy parameters we find that the cross-over
length can be as large as 10,000 bases. This is about the largest
size of naturally occurring RNAs as well as the largest length of RNA molecules
amenable to quantitative computational approaches. Thus, we conclude that
finite size effects have to be seriously taken into account when describing
RNA folding by asymptotic theories.

This article is organized as follows: In Sec.~II, we briefly review the definition 
of RNA secondary structure. In Sec.~III, we analytically derive the finite size effects 
of the simplest model of RNA folding namely a homogeneous sequence without loop entropies. 
While this model is mainly treated for pedagogical purposes, in Sec.~IV, we sketch how 
the result can be generalized to more realistic models of RNA folding. In Sec.~V, the behavior 
of the cross-over length $N_0$ is discussed. We find that $N_0$ depends mostly on the degree 
of branching of the RNA molecules and a simple approximate formula is derived. These results 
are shown to be consistent with the numerical values obtained using experimentally known energy 
parameters for specific sequences in Sec.~VI. We point out how enormous finite size effects in 
the RNA secondary structure formation problem can be. The detailed derivations of the partition 
function and the cross-over length are relegated to various appendices.

\section{Review of RNA secondary structures}

\subsection{Definitions}

RNA usually occurs as a single-stranded polymer with four types of monomers (bases) G, C, A, and U. The strand can bend back onto itself and form helices consisting of stacks of stable Watson-Crick pairs (A with U or G with C). 

\begin{figure}[h]
\includegraphics[width=8cm]{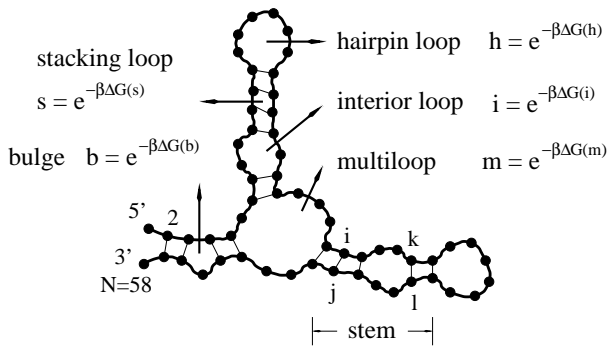}
\caption{An RNA secondary structure. The thick line stands for the backbone of the molecule and thin lines stand for base pairings. The solid dots represent monomers. 5' and 3' show the head and tail of this RNA of length 58. Many different loops formed when RNA folds are also defined in the figure.}
\label{SS}
\end{figure}

An RNA secondary structure describes which bases are bound and can be written as a set of binding pairs $(i,j)$, where $i$ and $j$ denote the $i^{th}$ and $j^{th}$ base of the RNA polymer respectively. For example, the secondary structure {\bf S} of Fig.~\ref{SS} is written as 
$$ S=\{(2,57),(3,56),...,(i,j),...,(k,l),(37,44)\}. $$

In this study, we apply the common approximation to exclude the so-called pseudo-knots~\cite{tinoco}, i.e., for two base pairs $(i,j)$ and $(k,l)$, the configurations $i \!<\! k \!<\! j \!<\! l$ and $k \!<\! i \!<\! l \!<\! j$ are not allowed. As a result, the analytical studies become more tractable. 

\begin{figure}[h]
\includegraphics[width=8cm]{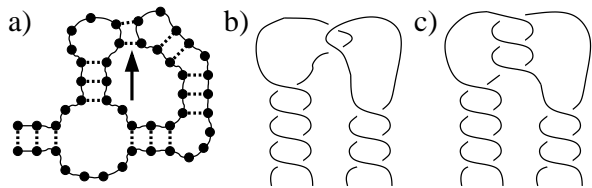}
\caption{Pseudo-knots in RNA structures: The base pairings indicated by the arrow in a) create a pseudo-knot. b) the short pseudo-knots (called ``kissing hairpins''). c) the long pseudo-knots in 3 dimensions.}  
\end{figure}

This exclusion of pseudo-knots is reasonable. For long pseudo-knots, the double helix structure would require threading one end of the molecule through its loops many times so they are kinetically difficult to form. Thus, these pseudo-knots occur infrequently in natural RNA structures~\cite{tinico,higgs00}. Short pseudo-knots, on the other hand, do not contribute much to the total free energy because of the relatively low binding energies for short sequences and the strong electrostatic repulsion of the backbone since the polymer backbone is highly charged. By excluding pseudo-knots, we will stay close to commonly used algorithms that compute the exact partition function which can be applied to test our model.

% find the statistics about the pseudoknots. 

\subsection{Interaction energies}

Since the tertiary interactions between structures are in general much weaker than the interactions among the secondary structures~\cite{tinoco,higgs00}, we will follow the common approaches and take into account only the energy contribution from the secondary structures.

If we assign a Gibbs free energy $\Delta G({\bf S})$ to each secondary structure {\bf S}, the partition function of the ensemble of all structures is given by
\begin{equation}
Z=\sum_{{\bf S}} e^{-\Delta G({\bf S})}.
\end{equation}

The Gibbs free energy is commonly used to describe the secondary structure since it contains entropic contributions from the formation of loops as well as enthalpic terms from the formation of base pairs. The total free energy $G({\bf S})$ is the sum of the energy contributed from each elementary piece such as the stacking of base pairs and the connecting loops. The largest contribution are the stacking energies between adjacent WC pairs, and these energies depend on the type of bases in the pairs. While the typical value of the stacking energy is on the order of $k_B T$ at room temperature, both the enthalpic and entropic terms are on the order of $10 k_B T$. Thus, the stacking energy will become repulsive with a moderate increase of temperature to around $80^{\circ}C$ and the RNA molecule denatures.

\section{Molten Phase}

In order to get a qualitative understanding of finite size effects, we first follow previous works~\cite{higgs96,pagna,bund99,higgs00,bund02} and assume that the Gibbs free energy is the sum of the binding free energies $\epsilon_{ij}$ of each base pair in the structure,
\begin{equation}
\Delta G({\bf S}) = \sum_{(i,j)\in {\bf S}} \epsilon_{ij},
\end{equation}
and neglect the entropic energies due to the formation of loops for the rest of this section. 

The binding free energies $\epsilon_{ij}$ in this model are differences between the gain in chemical binding energy and the loss in the configurational entropy associated with the formation of the base pairs. Since both contributions are large and comparable, the realistic values of the $\epsilon_{ij}$ strongly depend on temperature. Since we do not describe spatial degrees of freedom in this model, it does not describe denaturation of the RNA molecule and we restrict ourselves from here on to a parameter regime where the majority of the bases is paired, i.e., where a significant fraction of the $\epsilon_{ij}$ is negative.

In order to obtain analytical insights into the finite size effects, we additionally assume that the binding free energy $\epsilon_{ij}$ is a constant $\epsilon_0$, independent of the identities of the bases. Thus, in our simplified model $\Delta G({\bf S})=\epsilon_0 \times |{\bf S}|$ where $|{\bf S}|$ stands for the number of pairs in {\bf S}. This simplified energy model serves as the basis of our study for the more realistic energy model. 

As it stands, this energy model and the more realistic energy model we will introduce later describe only homogeneous sequences. However, it has been argued that this energy model can be applied to random RNA sequences at high enough temperature when the disorder is sufficiently weak~\cite{bund02}. Under this weak disorder, there exist many structures with nearly degenerate energies and the corresponding scaling laws match the predictions of the simplified energy model. Only as the temperature is lowered, a strong disorder phase arises. This low temperature phase is characterized by a small number of distinct low-energy structures, and is referred to as the ``glass phase'' in analogy with studies of other disorder system. However, this glass phase is not within the scope of this article.

\begin{figure}
\includegraphics[width=8cm]{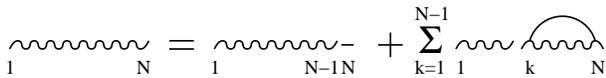}
\caption{Recursive relation for a simple model of an RNA molecule. The wavy lines stand for undetermined structure and the corresponding partition function. The arch represents a binding pair between bases k and N. The assumption of excluding pseudo-knots separates the last term into two independent parts since two pairs can not go across each other.} 
\label{molten}
\end{figure}

The partition function of the molten phase model can be obtained through the recursive relation in Fig.~\ref{molten}. This figure shows how the possible ways of binding can be decomposed into two cases where the last base N is either unbound or bound to some base k. If we define $G(N+1)$ as the partition function of an RNA of length N, the relation reads
\begin{equation}
G(N+1)=G(N)+\sum_{k=1}^{N-1} G(k)\cdot q \cdot G(N-k),
\end{equation}
where $q=e^{-\beta \epsilon_0}$. Together with the boundary condition $G(1)=1$, this equation allows calculation of the exact value of the partition function $G(N)$ in $O(N^2)$ time. The Vienna package~\cite{hofac} is able to calculate this exact value with more complete sequence dependent energy parameters based on the similar scheme. This recursive equation~(3) also leads us to the analytical expression for the partition function. By introducing the z-transform, 
\begin{equation}
\widehat{G}(z)=\sum_{N=1}^{\infty}G(N)\cdot z^{-N},
\end{equation}
and applying it to Eq.~(3), we get a quadratic equation for $\widehat{G}(z)$ as
\begin{equation} 
z\widehat{G}(z) - 1 = \widehat{G}(z) + q \widehat{G}^2(z),
\end{equation}
from which $\widehat{G}(z)$ can be solved. For large sequence lengths, the partition function $G(N)$ is obtained by performing the inverse z-transform on $\widehat{G}(z)$, and can be approximated (see Appendix A) as
\begin{eqnarray}
G(N) &=& \frac{1}{2\pi i}\oint \widehat{G}(z) z^{N-1} dz. \\ 
&\approx& A(q) N^{-\theta} z_c^N (q) \bigl[ 1-\frac{N_0(q)}{N}+O(\frac{1}{N^2}) \bigr],
\end{eqnarray}
where $z_c(q)=1+2\sqrt{q}$ is the branch point of $\widehat{G}(z)$, $\theta=3/2$, $A(q)=[(1+2\sqrt{q})/4\pi q^{3/2}]^{1/2}$ and $N_0(q)=3(1+4\sqrt{q})/16\sqrt{q}$. This asymptotic analytical formula is only determined by the behavior of $\widehat{G}(z)$ near the branch point $z_c$. The exponent $\theta=3/2$ indicates the characteristic universal behavior of this partition function for long sequences. The non-universal cross-over length $N_0(q)$ characterizes how long a sequence has to be for the universal laws to hold. Here, we find an explicit analytical formula for $N_0$ as a function of parameter q. 

From the formula of $N_0(q)$, we can see that the cross-over length in this simple model is on the order of 1 for all values of q. However, in the following sections we will show that the cross-over length may vary over several order of magnitudes from several bases to 1,000,000 bases. Thus, the loop entropies of the more realistic energy model would greatly modify the behavior of cross-over length.

\section{Including loop entropies} 

To get a more quantitative understanding of the cross-over length, we now take into account the loop entropies and introduce Boltzmann factors, $s$, $b$, $i$, $h$, and $m$, for the different types of loops (see Fig.~\ref{SS}). The values of these free energy parameters have been carefully measured~\cite{freier} such that our model can be applied quantitatively to realistic RNA molecules. Typically, the free energy of a stacking loop is large and negative ($s\gg1$) while the free energies for all the other loops tend to be large and positive, leading to Boltzmann factors much less than one. The binding energy $\epsilon_0$ of the simple model introduced above is now absorbed into these loop free energies. As mentioned in Sec.~III, we still restrict ourselves to a temperature regime below denaturation which is determined by the true energy model.

Again, we want to calculate the partition function of the structure ensemble and derive the cross-over length as a function of the loop parameters. This calculation in principle follows along the lines of Sec.~III, but is technically much more elaborate because of the more complicated energy model. A reader more interested in the final results than in the technical details is advised to directly skip to the Sec.~V. 

\begin{figure}[h]
\includegraphics[width=8cm]{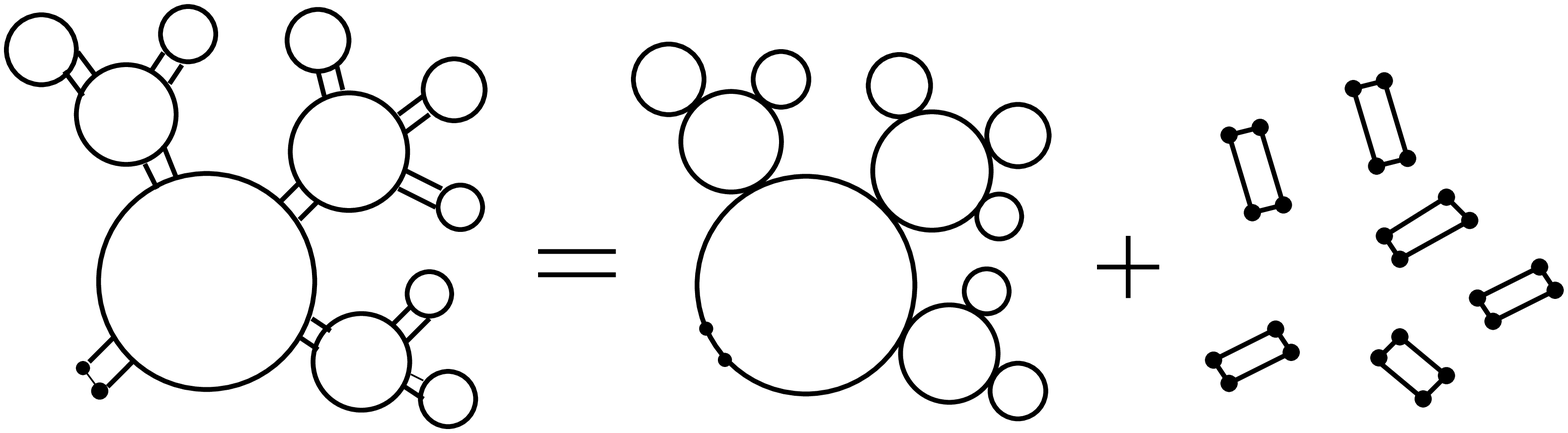}
\caption{Separation of stems from the bubble structure.}
\label{bubblestem}
\end{figure}

In order to calculate the partition function, we separate the secondary structure into two categories as shown in Fig.~\ref{bubblestem}. One is the bubble structure which contains only hairpins and multiloops. The other is the stem structure, which connects the bubbles, containing only stacking loops, bulges, and interior loops. We will study each of them individually and later combine them together. 

\subsection{Stem structure}

In principle, the loop free energy depends on the length of a loop. Thus, unbound bases also contribute to the total free energy. This contribution has been experimentally measured for small loops and behaves logarithmically with length when the loop is large. However, in the following we show that the free base energy of unbound bases provides only a negligible effect on the behavior of a stem and thus on the cross-over length if the stacking energy is relatively large. 

\begin{figure}[h]
\includegraphics[width=8cm]{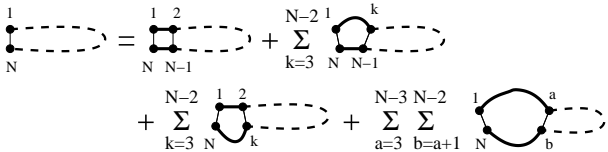}
\caption{Recursion relation for stem structure. The dashed lines stand for the undetermined structures. Thick lines represent the backbone and thin lines stand for pair bindings.} 
\label{stem}
\end{figure}

To explore all possible ways of pair bindings, again a graphical recursion relation is helpful. Such a recursion relation is shown in Fig.~\ref{stem}. Starting from a closed pair on the left, the following loop can be either a stacking loop, a bulge or an interior loop which correspond to the terms on the right hand side. To study the influence of a free energy for unbound bases, we assign the Boltzmann factors, $\tilde{b}$ and $\tilde{i}$, to each unbound base in a bulge and an interior loop respectively. If we define the partition function of stem structures with N bases and the first and the last bases of which is paired as $S(N-1)$, which corresponds to the left hand term, this relation in Fig.~\ref{stem} is formulated as
\begin{eqnarray}
S(N-1) &=& s S(N-3)+2 \sum_{k=3}^{N-2}b \tilde{b}^{N-k-1} S(k-2) \nonumber \\ 
&+& \sum_{a=3}^{N-3}\sum_{b=a+1}^{N-2}i \tilde{i}^{N-(b-a)-3} S(b-a).
\end{eqnarray}

To perform the z-transform, we have to consider the initial conditions, $S(1)=1$, $S(2)=0$, $S(3)=s$, and $S(4)=2b\tilde{b}$. These initial conditions arise because certain structures are not allowed, e.g., since a base can not be shared in two base pairings, a stem with 3 bases does not exist and this leads to $S(2)=0$. Also when the length of a stem is small, certain loops which require many bases are not allowed, e.g., the only available structure for a stem with 4 bases is a stacking loop. Including these conditions, we apply the z-transform on Eq.~(8) with the definition
\begin{equation}
\widehat{S}(z)=\sum_{N=1}^{\infty} S(N)z^{-N},
\end{equation}
to resolve the convolution. Solving for $\widehat{S}(z)$ gives us
\begin{equation} 
\widehat{S}(z) = \frac{1}{z} \left[ 1-\frac{s}{z^2}-\frac{2b\m \tb}{z^2(z-\tb)}-\frac{i\m \ti^2}{z^2(z-\ti)^2} \right]^{-1}.
\end{equation}

Again the partition function $S(N)$ can be obtained by applying the inverse z-transform on $\widehat{S}(z)$. 
 
To illustrate the effect from the unbound bases, we show how their Boltzmann factors affect the average behavior of the long stem structure. The average quantity of a certain type of loop or unbound base can be calculated as $\tau\cdot \partial_{\tau}\ln S(N)$ where $\tau$ is the corresponding Boltzmann factor. Here, we specifically calculate the average number of unbound bases per interior loop, which is defined by $\tilde{i}\cdot \partial_{\tilde{i}}\ln S(N) / i\cdot \partial_{i}\ln S(N)$ in the large N limit, as a function of $\tilde{i}$ (see Appendix B for detailed calculations).  

\begin{center}
\begin{figure}[h]
\includegraphics[width=7cm]{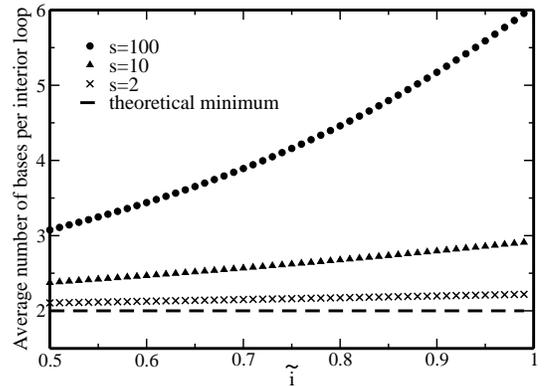}
\caption{Average number of unbound bases per interior loop v.s. $\tilde{i}$ for different stacking energy. Following the measured free energies~\cite{mathews}, we chose the typical values $b=0.01$, $\tilde{b}=0.85$, and $i=0.05$ for the other Boltzmann factors.}  
\label{avelenint}  
\end{figure}
\end{center}

From Fig.~\ref{avelenint}, we can see that this average number is barely affected by $\tilde{i}$ when the Boltzmann factor $s$ for a stacking loops is large. This can be easily understood as follows: For an interior loop, the unbound bases always introduce a strong energy penalty since less bases are available for stacking. This penalty is much larger than the penalty due to loss of the degree of freedom by one more free base. Thus when the binding energy is large, the free base entropic penalty can be neglected. From Fig.~\ref{avelenint}, we can see that interior loops tend to stay at the smallest length (2 unbound bases) when $s$ is large, independent of the value of $\tilde{i}$. Since the same argument applies to bulges as well, we will set $\tilde{i}$ and $\tilde{b}$ to 1 for the rest of this publication.

\subsection{Bubble structure}

\begin{figure}[h]
\includegraphics[width=8cm]{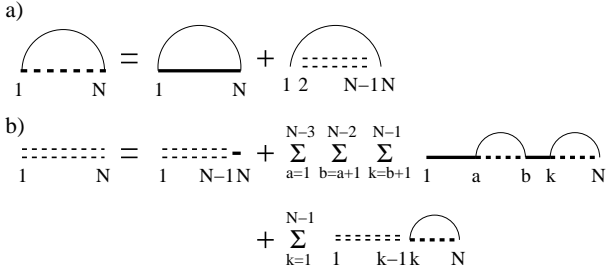}
\caption{Recursion relation for bubble structures. Dashed lines stand for undetermined structures. Thick lines and thin lines stand for the backbone and the pair binding, respectively. In a), the left hand term represents an undetermined bubble structure with the two end bases paired. The double dashed line stands for an undetermined multiloop structure and it can be decomposed into the components in b), as explained in the main text.}
\label{bubble}
\end{figure}

In a similar fashion, a recursive relation for the bubble structure is found graphically as shown in Fig.~\ref{bubble}. In the first relation for a closed bubble structure, we can have either a hairpin loop or a multiloop following from the closed pair at the end. In the second relation, the multiloop structure can be decomposed into two cases where the last base is either unbound or bound. Since a multiloop has to have at least 3 branches, we have a term with two more bubble structures; the last recursive term produces more branches. 

By defining the partition function for the closed bubble structure and multiloop structure with N bases as $B(N-1)$ and $M(N+1)$ respectively, the recursive relations read 
\begin{eqnarray}
B(N-1) &=& t \left( h + m\m M(N-1) \right) \\
M(N+1) &=& M(N) + \sum_{k=1}^{N-1} M(k)\m B(N-k) \\ \nonumber
       &+& \sum_{a=1}^{N-3} \sum_{b=a+1}^{N-2}\sum_{k=b+1}^{N-1} B(b-a) B(N-k).
\end{eqnarray}  

Here an additional Boltzmann factor $t$ is introduced in the first relation at the position where two bubbles are connected. Later we will insert stems into the bubble structure by replacing $t$ with the partition function of the stem structure. We also neglect the free base energy for unbound bases in hairpins and multiloops for similar arguments as above. 

In this recursive relation, the smallest multioop should have at least 4 bases such that two branchings can be connected. Thus, we set the initial conditions as $M(1)=M(2)=M(3)=M(4)=0$, which forbids a multiloop with length less than 4. With the initial conditions, the recursive relations result in the following quadratic equation for the z-transformed $\widehat{B}(z)$ :
\begin{equation}
\left( \frac{1}{t}+\frac{m}{(z-1)^2} \right) \widehat{B}^2 - \left( \frac{z-1}{t}+\frac{h}{z-1} \right) \widehat{B} + h = 0.
\end{equation}

\subsection{Complete structure}

\begin{figure}[h]
\includegraphics[width=8cm]{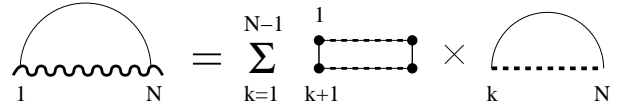}
\caption{Replacing the position of $t$ by stem structures. The left hand term represents the closed structure with both stem and bubble structures in it.}
\label{insert}
\end{figure}

To combine the stem and bubble structures, we insert a stem structure at each position represented by $t$ which is a placeholder for the connections between multiloops and hairpin loops in the bubble structure. In this case the first relation in Fig.~(\ref{bubble}) is modified as indicated in Fig.~\ref{insert}.

By defining the partition function of the closed structure on the left hand side as $C(N-1)$, the Fig.~\ref{insert} reads
\begin{equation}
C(N-1) = \sum_{k=1}^{N-1} S(k)\m B(N-k).
\end{equation}

After z-transform, this relation results in $\widehat{C}=(z\widehat{S})\widehat{B}$. Thus, the following replacement 
\begin{equation}
t \rightarrow z\widehat{S}, \qquad \widehat{B}\rightarrow \widehat{C}
\end{equation}
combines stem and bubble structures together. In order to complete all possible structures, the single strands outside the closed end pair also have to be included. This can be done by going back to the first recursion relation, Eq.~(3), and replacing $G(N-k)$ by the closed structures $C(N-k)$, which relates $\widehat{G}(z)$ to $\widehat{C}(z)$ as 
\begin{equation}
\widehat{G}=\frac{1}{(z-1)-\widehat{C}}.
\end{equation}

Putting everything together, we obtain 
\begin{eqnarray}
\widehat{G} &=& \frac{1}{2m} \Biggl[ \left( \frac{z-1}{z\widehat{S}}+\frac{2m-h}{z-1} \right) \\
&-& \sqrt{ \left( \frac{z-1}{z\widehat{S}}+\frac{h}{z-1} \right)^2 - \left( \frac{4h}{z\widehat{S}}+\frac{4hm}{(z-1)^2} \right)} \Biggr]. \nonumber
\end{eqnarray}

Notice that the leading singularity in $\widehat{G}$ is again from the branch cut induced by the square root. Thus, as expected, the inverse z-transform leads to the same universal behavior~(7) with an exponent $\theta=3/2$ as the simple model we studied first. However, non-universal quantities such as the cross-over length $N_0$ will depend on the parameters of the extended model.

\subsection{Minimum hairpin length constraint}

For natural RNA molecules, a hairpin loop needs to have at least 3 unbound bases due to the width of the double helix (which implies $j-i \!>\! 3$ in the secondary structure). This minimum hairpin length constraint is easily taken into account in our calculations. Under the constraint, a bubble structure which contains at least one hairpin must have at least 5 bases. Thus, we adopt the initial condition, $B(1)=B(2)=B(3)=0$, when we perform the z-transform on Eq.~(11). The summation range in Eq.~(11) is then changed and it simply leads to a substitution of $h$ by $h/z^3$ in all subsequent equations. 

This substitution is reasonable since $z$ represents the Boltzmann factor for the free energy of one single base. The minimum hairpin length constraint reduces the number of available bases for binding by 3, so it introduces an energy penalty $3\m \ln z$ which causes the Boltzmann factor $h$ to be divided by $z^3$. In this way, we can easily introduce any kind of minimum length constraint via a similar substitution. For example, if we require a bulge loop to have at least two unbound bases instead of one, the replacement of $b$ by $b/z$ will include this constraint.

Note, this principle also helps us to understand equation (10) for the stem structure. The terms with different powers of z, namely $s/z^2$, $b/z^3$, and $i/z^4$, arise because a stacking loop reduces the number of available bases for pairing by 2, a bulge loop has at least one unbound base, and an interior loop has a minimum of 2 unbound bases, respectively.

\section{Behavior of the cross-over length}

In this section, we will use the general results of Sec.~IV in order to calculate the cross-over length in our model for sequences with loop entropies.

\subsection{Large stacking energy approximation}

To derive the cross-over length, we need to solve for the branch point $z_c$ that is defined by the vanishing of the term under the square root in Eq.~(17). In principle, we can always obtain the numerical value of the branch point and expand $\widehat{G}(z)$ around this point to obtain a numerical value for the cross-over length (see Appendix A). We will refer to this numerical value as the homogeneous cross-over length since it is obtained under the homogeneous molten phase model.  
 
However, since this calculation involves finding the root of a fourth order polynomial, no meaningful analytical expression can be found in general. Thus, we resort to a large $s$ approximation in order to obtain an analytical expression. This approximation is justified since the Boltzmann factor of a stacking loop, $s$, is usually much larger than 1 while the loop Boltzmann factors $b,i,h,$ and $m$ are less than one. In this approximation, from Eq.~(17) we find the branch point $z_c\approx \sqrt{s}$, i.e., the free energy per base is $f=-kT\cdot \ln (z_c)\approx \frac{1}{2}\Delta G(s)$. This can be easily interpreted since we expect most bases to form pairs due to the favorable stacking loops such that the free energy per base is half of the free energy of a stacking loop.

\subsection{Cross-over length}

Including the minimum hairpin length constraint introduced in Sec.~IV.D, we expand the branch point $z_c$ near $\sqrt{s}$. Then, the approximated analytical formula for the cross-over length becomes
\begin{eqnarray}
N_0 &=& \frac{3s^{3/4}}{8\sqrt{hm}} \Big[ (\sqrt{s}-1)^2 + h \nonumber \\ 
  &+& \frac{9b}{2\sqrt{s}} + \frac{11i-6b}{4s} + O(s^{-3/2}) \Big]. 
\end{eqnarray}

It has a straight forward interpretation: The simplest possible structure is a long stem with one hairpin at the end. Every additional branching of the structure requires formation of one hairpin and one multiloop. Since upon formation of the hairpin and mulitloop, at least 3 bases become unbound, the Boltzmann factor for a branching is $hm/s^{3/2}$. The prefactor in Eq.~(18) is up to the numerical factor of $3/8$ the inverse square root of this expression. Thus, we conclude that the cross-over length which becomes larger as the Boltzmann factor for a single branching becomes smaller can be interpreted as the minimum length that allows a certain degree of branchings.  
   
Since the Boltzmann factors $b$ and $i$ appear only in the higher order terms, they barely modify the cross-over length and can be neglected altogether. This is consistent with the fact that $b$ and $i$ only play roles in the stem structure, but not in the bubble structure. The leading term of the approximated analytical formula~(18) will be referred to as the large s cross-over length. 

\subsection{Reliability of the large s approximation}

For the analytical large s cross-over length to be useful, we have to know how good they agree with the numerical value of the homogeneous cross-over length. Here, we compare these two values for many different choices of energy parameters covering the whole range of realistic values. Fig.~9 shows how the large s cross-over length approaches the homogeneous one as $s$ gets large. Typical values for the Boltzmann factor of a stacking loop $s$ involving GC pairs are $s\geq 30$~\cite{freier}, so the approximation is very good in this region. For stacking loops involving AU pairs, $s$ is around 5 so a deviation from the approximated formula in the large $s$ limit can be seen. However, since $N_0$ only sets the order of magnitude of the length beyond which the asymptotic theory is applicable, the large s cross-over length with a deviation by a factor of 2 at $s=5$ is still a good estimation. 
 
\begin{center}
\begin{figure}[h]
\includegraphics[width=7cm]{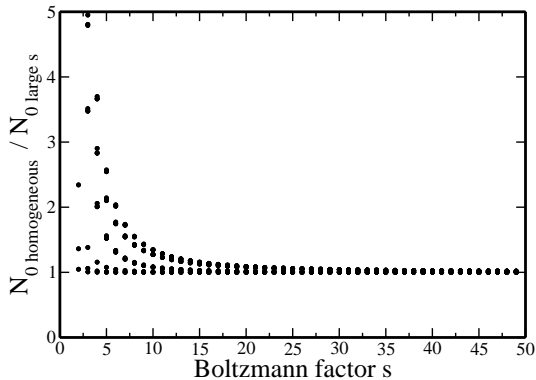}
\caption{Ratio between the large s cross-over length and the numerical value of the homogeneous one for many combinations of the parameters: $b,i=\{0,1\}$ and $h,m=\{0.1,0.01,0.001\}$. These choices cover the region of realistic values.}  
\label{N0s}
\end{figure}
\end{center}

\section{Numerical verification}

\begin{figure}[h]
\includegraphics[width=8cm]{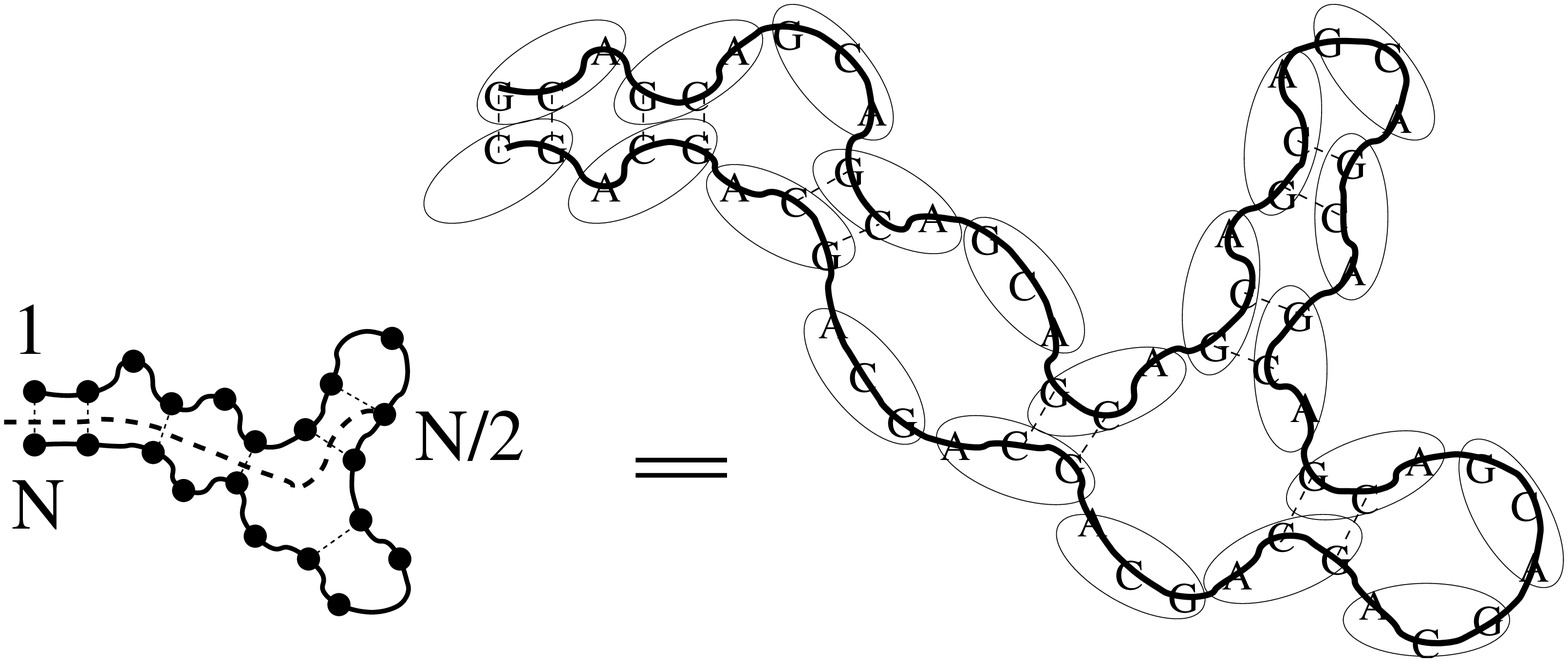}
\caption{Equivalence between the $(GCA)_n$ sequence and the molten phase : Three consequent bases GCA are mapped to one single base so the GC/CG stack is equivalent to a binding pair in the molten phase. Then, the smallest interior loop on the left is viewed as a stacking loop and the following interior loop becomes a bulge in the molten phase. The hairpin loop on the right with 2 units of GCA is considered to have two unbound bases in the molten phase.}
\label{GCA}
\end{figure}

While for a generic RNA sequence, the molten phase model is believed to only apply at sufficiently high temperature, it can be applied to repeated sequences at all temperatures below denaturation since each repeated unit can be viewed as the equivalent of a base in the molten phase (Fig.~\ref{GCA}). To illustrate the correctness of the calculations shown in Secs.~IV and V, and to get a feeling for typical cross-over lengths, we now compare our large s cross-over length of repeated sequences with the full numerical results. The full numerical result is obtained by using the Vienna package~\cite{hofac} which can calculate the exact value of the partition function and other observables for arbitrary sequences using a realistic energy model. 

As an observable, we choose the average size $l$ of a structure. This quantity is defined as 
\begin{equation}
l = \sum_{k=1}^{N/2} \sum_{k'=N/2+1}^{N} P_{k,k'},
\end{equation}
where $P_{k,k'}$ is the probability that bases $k$ and $k'$ are paired. The latter probability is also calculated exactly by the Vienna package. This size measures the average number of base pairs to be crossed when connecting the $N/2^{th}$ base to base 1 (Fig.~\ref{GCA}), which captures the size of the secondary structure. We expect $l$ to obey
\begin{equation}
l \propto N^{1/2}\cdot \left( 1-(\frac{N_0}{N})^{1/2}+O(\frac{1}{N}) \right),
\end{equation}
where the leading term is the asymptotic behavior~\cite{bund02} and the next term reflects the first expected correction which is a constant independent of N. We determine $l$ for sequences of different lengths and extract the full numerical cross-over length $N_0$ by fitting data obtained via the Vienna package to Eq.~(20). This is then compared to our large s cross-over length.

\subsection{$(GCA)_n$ sequence}

We apply this scheme to a repeated $(GCA)_n$ sequence. Such sequences naturally occur in the gene for Huntington's disease and their secondary structures are believed to play a role in this disease. Since the GC/CG stack in the secondary structure of the $(GCA)_n$ sequence is much more favorable than any other combination, we can exclude the possibility to have binding pairs other than GC/CG. Thus, GCA is viewed as one unit base in the molten phase. With this equivalence we can use the experimentally determined parameters~\cite{mathews} and calculate the equivalent energy parameters $s$, $b$, $i$, $h$, and $m$ for the molten phase model. For example, the stacking energy of the molten phase is the sum of GC/CG stacking energy and the free energy for the interior loop of length 2 in the $(GCA)_n$ sequence.

\begin{center}
\begin{figure}[h]
\includegraphics[width=7cm]{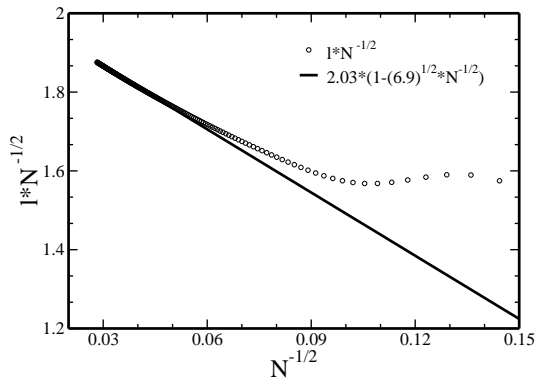}
\caption{Average size of the secondary structure of a $(GCA)_n$ sequence at $37^{\circ}C$. The data is fitted by the expected law~(20). }
\label{GCAfit}
\end{figure}
\end{center}

Fig.~\ref{GCAfit} shows the full numerical results for the average size of the secondary structure for a $(GCA)_n$ sequence as a function of the sequence length $N$. By fitting the result to the Eq.~(22), we get a full numerical value of the cross-over length of 6.9 bases. 

To compare this full numerical value with our large s cross-over length, we plug the equivalent energy parameters of the molten phase model into the approximated formula $3s^{1/4}(\sqrt{s}-1)^2 / 8\sqrt{hm}$. This formula is different from equation~(18) because the minimum hairpin length of the corresponding molten phase model only 1 instead of 3. The resulting large s cross-over length is about 2.3 repeat units which corresponds to 6.9 bases in the $(GCA)_n$ sequence. This agrees very well with the full numerical value.

\subsection{$(AU)_n$ sequences}

Repeated AU sequences have already been suggested as models for the molten phase by de Gennes in 1968~\cite{degen}. For such sequences, we exclude the possibility of AA or UU binding pairs since they are not favorable at all. In a similar fashion as for the $(GCA)_n$ sequence in Fig.\ref{GCA}, the smallest bulge has two free bases. However, since the minimum hairpin length is 4 instead of 3, in order to match AU or UA closing pairs, the large s cross-over length is $3s(\sqrt{s}-1)^2 / 8\sqrt{hm}$.

\begin{center}
\begin{figure}[h]
\includegraphics[width=7cm]{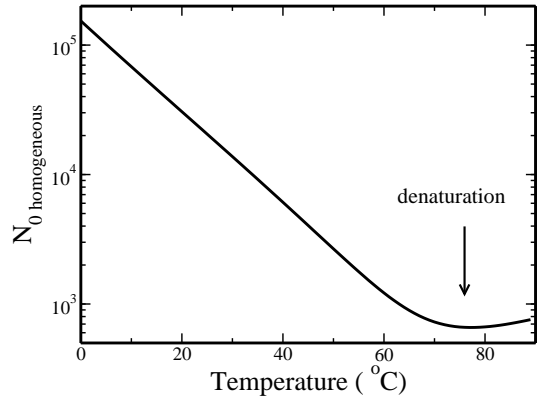}
\caption{Numerical values of homogeneous cross-over length for $(AU)_n$ sequence with respect to temperatures. This quantity is obtained by numerically calculating the branch point and then expanding $\widehat{G}(z)$ around this point(see Appendix A).} 
\label{N0AUT}
\end{figure}
\end{center}

Plugging in the correct values for the parameters at body temperature results in a large s cross-over length of $N_0\approx 7700$. A verification of this value is beyond the reach of the numerical procedure using the Vienna package. Since the cross-over length is expected to decrease as the denaturation transition is reached, the full numerical verification could be performed at a higher temperature. Thus, we first study the the temperature dependence of the cross-over length $N_0$. To this end we choose to study the homogeneous cross-over length instead of large s cross-over length because it is not clear whether the large s approximation is appropriate when approaching the denaturation temperature. When studying the temperature dependence, it is important to note that the free energies themselves have a very strong temperature dependence which changes Boltzmann factors much more than the explicit $\beta$ in the exponent. By looking up the table of enthalpies corresponding to different loops, we determine this temperature dependence of the Boltzmann factors $s$, $b$, $i$, $h$, and $m$. The numerical values of the homogeneous cross-over lengths thus obtained are shown in Fig.~\ref{N0AUT}. From this figure, we choose to numerically verify the estimate of the cross-over length at $57^{\circ} C$.  

The full numerical verification is done in a similar way as Sec.~VI.A. Fitting numerical data at $57^{\circ} C$ gives a full cross-over length $N_0 \approx 900$. Our homogeneous cross-over length predicts a value $N_0 \approx 1500$ which is of the same order of magnitude as the full numerical result.      
From the above numerical studies, we conclude that our theory for the large s cross-over length is a good estimate. Thus, the cross-over length of $(AU)_n$ at body temperature is truly around 7,700 bases, which is almost the largest size of naturally occurred RNA sequences. For $(GC)_n$ sequences at body temperature, the situation is even more dramatic with a predicted cross-over length about 105,000 bases. This is beyond the limit of most natural RNAs and suggests that the structure ensemble of $(AU)_n$ and $(GC)_n$ molecules can never be described by the asymptotic theory for any naturally available molecules. 

\subsection{Distribution of cross-over length for short repeated sequences}

Lastly, we want to explore the sequence dependence of the cross-over length for short repeated sequences. We study all possible partially self-complementary repeated sequences of unit length two and three and evaluate their cross-over lengths using the formula for the large s cross-over length. 

\begin{figure}[h]
\includegraphics[width=8cm]{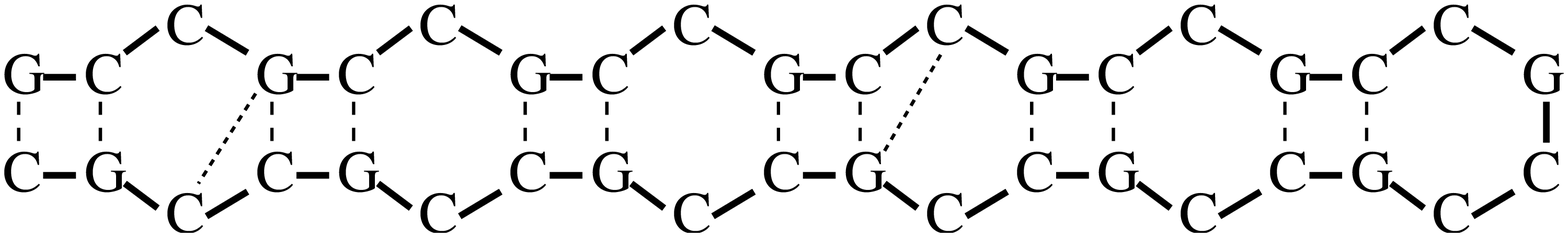}
\caption{A $(GCC)_n$ stem in the ground state. The dash lines stand for base pairing of ground state. The dot lines show the possibilities for base G to pair with neighboring C.}
\label{GCC}
\end{figure}

For non self-complementary repeated sequences, e.g., the $(GCC)_n$ sequence, it is not a priori clear that we can map the repeated unit GCC onto one single base in the molten phase (Fig.~\ref{GCC}) since base G can pair with either one of the two bases C. However, we verified numerically that more than 99\% of the G's are paired in the stacked configuration indicated by the dashed lines in Fig.~\ref{GCC}. This is due to the large loss in free energy as one stacking loop is lost and a bulge loop is created. We verified this behavior for all the partially self-complementary sequences in our discussion. Thus, we can safely use the same approach as used for the $(GCA)_n$ sequence and calculate the large s cross-over length. The results are shown in the following table.     

\begin{center}
\begin{table}[h]
\begin{tabular}{|l|c|c|c|c|c|c|c|c|}
\hline
Seq   & UAA  & UAC & GCA & CGA & GU & GCU & AUU & AUC \\ \hline 
$N_0$ & 4 & 7 & 7 & 34 & 40 & 70 & 115 & 130  \\ \hline
Seq   & GCC  & CGU & UAC & AUG & AU & CCG & GC & \\ \hline 
$N_0$ & 450 & 2800 & 3900 & 5600 & 7700 & 10700 & 105000 & \\ \hline
\end{tabular}
\caption{Large s cross-over length of short repeated sequences.}
\end{table}
\end{center}

We can see that the cross-over lengths spread over the whole range from only several bases to $10^6$ bases. This non universal behavior has to be treated individually for different sequences. For longer repeated units, we expect similar behavior but we did not verify this due to the complicated entropic effects. 

\section{Conclusion}

For our asymptotic description to be appropriate, an RNA sequence must be long enough to allow a certain degree of branching. The cross-over length at which this happens for homogeneous sequences is strongly sequence dependent. Specifically for cases when stacking is very favorable and hairpins and multiloops are unfavorable, the cross-over length $N_0$ can be very large. In this region, finite size effects are very important and have to be taken into account in any numerical simulation or interpretation of experimental data in terms of asymptotic formulas. Our analytical results provide an easy way to estimate this cross-over length for realistic energy models.

It remains as an interesting and relevant question how large the cross-over lengths for generic, non-homogeneous sequences in the glassy phase are. Unfortunately, this question can at this stage neither be addressed numerically nor analytically since even the expected asymptotic behavior is elusive. Nevertheless, the existence of large cross-over length in the molten phase as established here suggests that also in the glass phase large cross-over length have to be considered as a serious possibility.

\section{Appendix}

\subsection{z-transform}

For a series of numbers \{G(1), G(2), ..., G(N), ...\}, the z-transform of G(N) is defined as
$$ \widehat{G}(z)=\sum_{N=1}^{\infty}G(N)\m z^{-N}.$$

To recover $G(N)$, we apply the inverse z-transform:
$$ G(N)=\frac{1}{2\pi i} \oint_C G(z)\m z^{N-1}dz. $$

As an example, here we solve for the partition function for the molten phase model. From Eq.~(5), $\widehat{G}(z)$ is calculated as 
$$ \widehat{G}(z)=\frac{1}{2q}(z-1-\sqrt{(z-1)^2-4q}), $$
so the partition function $G(N)$ can be obtained by plugging $\widehat{G}(z)$ into the inverse z-transform.
 
The analytical part $(z-1)$ can be dropped since it does not contribute to the integral. The square root introduces a branch cut with two branch points at the ends. By taking the contour C as the smallest loop around the branch cut, the integral becomes 
\begin{eqnarray*}
G(N) &\approx& \frac{1}{4\pi qi} \int_{z_0}^{z_c} \Bigl( \sqrt{(z-1)^2 -4q+i\epsilon}\\
&\qquad& -\sqrt{(z-1)^2-4q-i\epsilon} \Bigr)\m z^{N-1}dz \\
&=& \frac{1}{2\pi q} \int_{z_0}^{z_c} \sqrt{4q-(z-1)^2}\m  z^{N-1}dz,
\end{eqnarray*}
where $z_c=1+2\sqrt{q}$ is the branch point with greatest real part and $z_0$ is the other branch point. In the large N limit, due to the exponent factor $z^N$, we expect that only the behavior near $z_c$ is important, so we can expand the integrand around $z_c$ and perform the approximation. 

To do this integral, we first consider a more general case where $\widehat{G}(z)=\sqrt{f(z)}$. The behavior near the greatest value of $z$ is obtained correctly if we replace $z$ by $e^{\mu}$ and expand $\sqrt{f(\mu)}$ in the power terms of $(\mu_c-\mu)$ as 
\begin{eqnarray*}
\sqrt{f(\mu)} &=& \sqrt{-f'(\mu_c)} (\mu_c-\mu)^{1/2} \\
&\qquad& \cdot \Bigl[ 1 - \frac{f''(\mu_c)}{4f'(\mu_c)}\m (\mu_c-\mu) + O((\mu_c-\mu)^2) \Bigr].
\end{eqnarray*}

Together with the following approximation valid in the large N limit, 
$$ \int_{\mu_0}^{\mu_c} (\mu_c-\mu)^{\alpha}e^{\mu N}d\mu \approx \Gamma(1+\alpha)N^{-(1+\alpha)}e^{\mu_cN}, $$
the inverse z-transform of $\widehat{G}(z)$ can be expressed as 
\begin{eqnarray*}
G(N) &=& \frac{1}{\pi}\int_{\mu_0}^{\mu_c} \sqrt{f(\mu)} e^{\mu N} d\mu \\
&=& \frac{\sqrt{-f'(\mu_c)}}{\pi} \Gamma(\frac{3}{2}) N^{-3/2}e^{\mu_cN} \\
&\qquad& \cdot \Bigl[ 1 - \frac{f''(\mu_c)}{4f'(\mu_c)}\m \frac{\Gamma(5/2)}{\Gamma(3/2)}\m \frac{1}{N} + O(\frac{1}{N^2}) \Bigr] \\
&=& A_0 N^{-3/2} z_c^N \Bigl[ 1 - \frac{N_0}{N} + O(\frac{1}{N^2}) \Bigr].
\end{eqnarray*}
Here, the partition function of the molten phase model~(7) is obtained by setting $f(\mu)=4q-(e^{\mu}-1)^2$. For the general case, we can put the result in terms of $z$ with the substitutions
\begin{eqnarray*}
\frac{df(\mu)}{d\mu}\Bigr|_{\mu_c} &=& z_c\m \frac{df(z)}{dz}\Bigr|_{z_c} \\
\frac{d^2f(\mu)}{d\mu^2}\Bigr|_{\mu_c} &=& z_c\m \frac{df(z)}{dz}\Bigr|_{z_c} + z_c^2\m \frac{d^2f(z)}{dz^2}\Bigr|_{z_c}. 
\end{eqnarray*}
After little algebra, the cross-over length is obtained as 
$$ N_0 = \frac{3}{8}\m \Bigl[1+\frac{z_c\m \frac{d^2f(z)}{dz^2}\bigr|_{z_c}}{\frac{df(z)}{dz}\bigr|_{z_c}} \Bigr] $$
for arbitrary $f(z)$.  

\subsection{average length of interior loops}

The calculation of partition function for the stem structure $S(N)$ is different from that for the molten phase model because there is no branch cut, so we do not expect the universal behavior $N^{-3/2}$. However, the fact that only the pole with the greatest real part is important in the large N limit is preserved. From Eq.~(8), the z-transform of the partition function for the stem structure $\widehat{S}(z)$ has the form $f(z)/g(z)$ where 
\begin{eqnarray*}
g(z) &=& z^2(z-\tilde{b})(z-\tilde{i})^2 - s(z-\tilde{b})(z-\tilde{i})^2 \\ 
&\qquad& - 2b\tilde{b}(z-\tilde{i})^2 - i\tilde{i}^2 (z-\tilde{b}). 
\end{eqnarray*}
In the large N limit, the inverse z-transform can be approximated by 
\begin{eqnarray*}
S(N) &=& \frac{1}{2\pi i} \oint \widehat{S}(z) z^{N-1}dz \\
&\approx& \frac{1}{2\pi i} \oint \frac{f(z)}{g'(z_c)(z-z_c)} z^{N-1} dz \\
&\approx& \frac{f(z_c)}{z_c g'(z_c)} z_c^N,
\end{eqnarray*}
where $z_c=z_c(s,b,\tilde{b},i,\tilde{i})$ is the pole with greatest real part. 

Since $z_c^N$ is the dominating term for large N, the average quantity can be approximated as
\begin{eqnarray*}
\langle \tau \rangle &=& \tau \partial_{\tau} \ln(S(N)) \\ 
&\approx& N \tau \partial_{\tau} \ln(z_c) = N z_c^{-1} \tau \partial_{\tau} z_c.  
\end{eqnarray*}
So the average number of unbound bases per interior loop can be obtained by 
$ (\tilde{i} \partial_{\tilde{i}} z_c) / (i \partial_i z_c) $.
Here, the pole $z_c$ can only be solved numerically.

\end{document}